\documentclass{article}
\usepackage{graphicx}
\usepackage{epsfig}
\usepackage{amsfonts}
\usepackage{enumerate} 
 \usepackage{amssymb}
\usepackage{cite}
\usepackage{enumitem}

 \usepackage{amssymb}

\date{\today} 

\newcommand{\ee}{\end{equation}}
\newcommand{\eea}{\end{eqnarray}}
\newcommand{\be}{\begin{equation}}
\newcommand{\bea}{\begin{eqnarray}}

\begin{document}

\title{\bf  Static black holes without spatial isometries:
\\
from AdS multipoles to electrovacuum scalarization }

\author{
{\large Carlos Herdeiro}
and
{\large Eugen Radu}
\\ 
\\
{\small Departamento de Matem\'atica da Universidade de Aveiro and} \\
{\small  Center for Research and Development in Mathematics and Applications (CIDMA)} \\ 
{\small Campus de Santiago, 3810-183 Aveiro, Portugal}
}
\date{May 2020}
\maketitle

\begin{abstract}   
We review recent results on the existence of static black holes without spatial isometries
in four spacetime dimensions
and propose a general framework for their study.
These configurations  are regular on
and outside a horizon of spherical topology. 
Two different mechanisms allowing for their existence are identified.
The first one relies on the presence of a solitonic limit of the black holes; when the solitons have no spatial isometries, the black holes, being a non-linear bound state between the solitons and a horizon, inherit this property. 
The second one is related to black hole scalarization, and the existence of zero modes of the scalar field without isometries around a spherical horizon.  When the zero modes have no spatial isometries, the backreaction of their non-linear continuation makes the scalarized black holes inherit the absence of spatial continuous symmetries.
 A number of  general features of the solutions are discussed together with possible 
generalizations.
\end{abstract}

 \tableofcontents

\section{Introduction} 

An interesting property of some  non-linear  field theory models
in a flat spacetime background
 is the existence of solitonic 
configurations\footnote{By solitons we mean non-singular,
finite energy localized solutions, without any assumption about their stability.} 
which are static but not spherically symmetric 
\cite{Manton:2004tk}.
Moreover, in some cases, the configurations minimizing the action in a given sector of the theory
possess discrete symmetries only. Examples include 
magnetic monopoles in Yang-Mills-Higgs models~\cite{Houghton:1995bs}, Hopfions~\cite{Faddeev:1996zj,Battye:1998zn} and Skyrmions 
\cite{Battye:1997qq}.
 
Rather surprisingly, the situation in Einstein's theory of gravity is different\footnote{In this work we shall restrict to
the case of a four dimensional spacetime.}.  The natural counterparts of the aforementioned field theory solitons, 
as non-perturbative elementary solutions, are black holes (BHs)\footnote{We recall that, for a Minkowski spacetime background, no smooth
horizonless configurations exist in Einstein's theory 
\cite{Lichnerowicz} -  see $e.g.$ the discussion in~\cite{Herdeiro:2019oqp}.
}. 
However, as implied by a number of 
classic results \cite{Chrusciel:2012jk}, 
the spectrum of BHs is very limited;
at least in the (electro)vacuum case
the solutions possess a high degree of symmetry.
For example, Israel's theorem
\cite{Israel:1967za,Israel:1967wq},
has established that a static BH solution in  Einstein-Maxwell theory
is spherically symmetric 
and described by only two parameters, its ADM mass and electric/magnetic charge. 

\medskip

The extrapolation of these results for a more general matter content
and/or different spacetime asymptotics is, however, unjustified.
In particular, in analogy with the situation in soliton physics,
in certain models one finds static BH solutions {\it without}
any continuous spatial symmetry.
 
To the best of our knowledge, two different mechanisms allowing for  
such configurations have been identified in the literature.
The first one can easily be 
understood from the discussion above and
relies on three different ingredients:
 \begin{enumerate}[label=\roman*)]
\item
the existence of (non-gravitating) solitons in certain field theory models,
which are static and non-spherically symmetric;
\item
these configurations possess generalizations as test fields on a Schwarzschild BH background\footnote{This step is nontrivial: not all solitons
survive when adding a BH horizon at their center.
For instance, spherically symmetric boson stars admit no BH generalization \cite{Pena:1997cy}.},
$i.e.$
one can add a (small) horizon at the center of a soliton \cite{Kastor:1992qy};
\item
the  solutions  survive in the fully non-linear regime, when including the soliton's backreaction\footnote{This is not
automatically guaranteed. For example, the BPST Yang-Mills solution is destroyed
when including gravity effects, for an asymptotically flat, five-dimensional spacetime
\cite{Volkov:2001tb}.
}, without any enhancement of symmetry.
\end{enumerate} 
Restricting to the asymptotically flat case, 
a number of partial results in this direction can be found in Refs.
\cite{Ridgway:1995ke}-\cite{Kleihaus:2013tba}.
%
%
However, all these  configurations still possess a Killing vector associated with rotational symmetry.
Although BHs without isometries
should also exist in those cases
no explicit  non-perturbative construction has been reported,
presumably due to the highly non-linear considered matter models
(see, however, the partial results in 
\cite{Ioannidou:2006mg}).

The work reported in \cite{Herdeiro:2016plq} has provided 
an explicit realization of this mechanism,
the technical obstacle of dealing with non-linear matter fields being avoided
by considering an electromagnetic field and AdS asymptotics.
Then, steps i)-iii) above are easily fulfilled, which results in a tower of solutions
without isometries,
 naturally interpreted as Schwarzschild-AdS BHs with electric (or magnetic)
multipole hair.

\medskip
The second mechanism 
allowing for static BHs with discrete spatial symmetries only
is rather different, and relies on the phenomenon of 
{\it spontaneous
scalarization}\footnote{Similar solutions should also exist for higher spin 
fields  ($e.g.$ spontaneous
vectorization), in which case, however,
the numerical problem becomes more complicated.
}.
Two different steps can be identified in this case:
 \begin{enumerate}[label=\roman*)]
\item
the static, spherically symmetric BHs $S_0$ of a given model
are
unstable against scalar perturbations, with the existence of {\it scalar clouds}.
That is, the model possesses zero modes, with an infinitesimally small scalar
field, which, however, exist for a certain set of $S_0$-parameters.  
\item
The scalar clouds survive when including backreaction
on the spacetime geometry,
with a bifurcation to  a new family
 $S_e$ of BH solutions.
In particular, there are zero modes 
 resulting in $S_e$-BHs with no
rotational symmetry at all.
\end{enumerate}
An explicit realization of these setup has been investigated in the recent work
\cite{Herdeiro:2018wub},
which has studied the
spontaneous scalarization of Reissner-Nordstr\"om (RN) BH.
As discussed there,
unlike the case of electrovacuum, the Einstein-Maxwell-scalar model admits
static, asymptotically 
flat, regular on and outside the horizon BHs without any spatial isometries.

\medskip
 
 This paper is structured as follows.
In the next Section we exhibit a general framework for the study of static BHs with
no rotational symmetries. 
Sections \ref{sec_1} and \ref{sec_2} discuss explicit realizations of the two
mechanisms discuss above, for Einstein-Maxwell-AdS and  Einstein-Maxwell-scalar theories,
respectively.
 Finally, in Section~\ref{sec_3} we present some conclusions and further remarks.
The Appendix contains technical details on the far field asympototics of
Einstein-Maxwell-AdS BHs

\section{Static black holes with no isometries:
a general framework}
 \label{sec_0}

\subsection{The metric} 

In the absence of  closed form solutions, static BHs without spatial isometries are constructed numerically.
This
task, however, is a numerical challenge,
since the geometry, and the matter functions, depend on all three space coordinates, making
the expression of the Einstein tensor very complicated.
Also, some techniques which work well for axially symmetric General Relativity 
problems 
($e.g.$ the metric gauge choice)
do not have a straightforward extension in this case.

These difficulties can be 
(at least partially) circumvented 
by employing the Einstein-De Turck (EDT) approach, proposed in~\cite{Headrick:2009pv,Adam:2011dn}. 
This approach has become, in recent years, a standard tool in the numerical treatment of 
stationary problems in general relativity, 
and has the advantage of not fixing \textit{a priori} 
a metric gauge, 
yielding at the same time elliptic equations -
 see~\cite{Wiseman:2011by,Dias:2015nua} for reviews.    

In this  approach  one solves the so called EDT equations 
\begin{eqnarray}
\label{EDT}
R_{\mu\nu}-\nabla_{(\mu}\xi_{\nu)}= 8\pi G  \left(T_{\mu\nu}-\frac{1}{2}T  g_{\mu\nu}\right) \ .
\end{eqnarray}
Here, $\xi^\mu$ is a vector defined as
$
\xi^\mu\equiv g^{\nu\rho}(\Gamma_{\nu\rho}^\mu-\bar \Gamma_{\nu\rho}^\mu)\ ,
$
where 
$\Gamma_{\nu\rho}^\mu$ is the Levi-Civita connection associated to the
spacetime metric $g$ that one wants to determine, and a reference metric $\bar g$ is introduced, 
$\bar \Gamma_{\nu\rho}^\mu$ being the corresponding Levi-Civita connection;  
$T_{\mu\nu}$ is the energy-momentum tensor of the matter field(s) and $G$
is Newton's constant.

Solutions to (\ref{EDT}) solve the Einstein equations
iff $\xi^\mu \equiv 0$ everywhere.
To achieve this,
we impose boundary conditions  which are compatible with
$\xi^\mu = 0$
on the boundary of the domain of integration.
Then, this should imply $\xi^\mu \equiv 0$ everywhere,
a condition which is verified from  the numerical output. 
 
  The BHs with no isometries
are more easily  constructed by employing 
spherical coordinates $(r,\theta,\varphi)$,
such that the horizon is located at some constant value of $r$.
As such, one introduces a general metric ansatz
with seven functions, $F_1,F_2,F_3,F_0,S_1,S_2,S_3$:
\begin{eqnarray}
\label{metric}
&&
ds^2=-F_0(r,\theta,\varphi) N(r)dt^2+F_1(r,\theta,\varphi) \frac{dr^2}{N(r)}+F_2(r,\theta,\varphi)\left(r d\theta+S_1(r,\theta,\varphi) dr \right)^2
\\
\nonumber
&&
\ \ \ \ \ \ \ \  +F_3(r,\theta,\varphi)  \big(r \sin \theta d\varphi+S_2(r,\theta,\varphi) dr+S_3(r,\theta,\varphi) r d\theta \big)^2
\ .
\end{eqnarray}
 $N(r)$ is an input 'background' function
whose explicit form depends on the considered problem. 
A general enough expression for  $N(r)$ which covers both models in this paper 
 is  
\begin{eqnarray}
\label{N}
N(r)=\left(1-\frac{r_H}{r}\right)\left(1+\frac{r^2+r_H^2}{L^2}+\frac{r r_H}{L^2}-\frac{q^2}{r r_H}\right),
\end{eqnarray}
with $r_H\geqslant 0$ denoting the event horizon radius 
and $q,L$ two input constants which will be defined below.
The reference metric $\bar g$ is found by taking 
$F_1=F_2=F_3=F_0=1$,
$S_1=S_2=S_3=0$ in the ansatz (\ref{metric}).

We remark  that the static axially symmetric  (or spherically symmetric) BHs can also be studied within framework.
For example, the  axially symmetric metric has
\begin{eqnarray}
\label{ax-lim}
 S_2=S_3=0,
\end{eqnarray}
all remaining functions depending on $(r,\theta)$ only.

\subsection{The boundary conditions} 

In this approach, the problem reduces to
solving a set of seven partial differential equations (PDEs) resulting from (\ref{EDT})-(\ref{metric}),
for the metric functions,  
together with a set of PDEs for the matter functions.
These equations are solved with suitable
boundary conditions (BCs)
which are found by  constructing an approximate form of the solutions on the
boundary of the domain of integration compatible with the requirement $\xi^\mu = 0 $ 
plus
regularity and AdS/Minkowski asymptotics of the solutions, as appropriate for the problem at hand.

 The configurations reported in this work 
possess
a reflection symmetry along the equatorial plane
($\theta=\pi/2$)
and two $\mathbb{Z}_2$-symmetries $w.r.t.$ the $\varphi-$coordinate.
Then the domain of integration  for the $(\theta,\varphi)$-coordinates 
is $[0,\pi/2]\times [0,\pi/2]$.

The metric functions satisfy the following
 BCs at infinity:
\begin{eqnarray}
\label{infinity}
 F_1= F_2=  F_3= F_0=1,~
S_1=S_2= S_3=0,
\end{eqnarray}
such that the background geometry is approached.
 The BCs at $\theta=0$ are
\begin{eqnarray}
\label{t0}
\partial_\theta F_1= \partial_\theta F_2=  \partial_\theta F_3= \partial_\theta F_0=0,~
S_1=S_2= \partial_\theta S_3=0, 
\end{eqnarray}
while at $\theta=\pi/2$ we impose
\begin{eqnarray}
\label{tpi2}
\partial_\theta F_1= \partial_\theta F_2=  \partial_\theta F_3= \partial_\theta F_0=0,~
S_1=\partial_\theta  S_2=   S_3=0.
\end{eqnarray}
The BCs at $\varphi=0$ are
\begin{eqnarray}
\label{fi0}
\partial_\varphi F_1= \partial_\varphi F_2=  \partial_\varphi F_3= \partial_\varphi F_0=0,~
\partial_\varphi S_1=   S_2=   S_3=0, 
\end{eqnarray}
while at
 $\varphi=\pi/2$ we impose
\begin{eqnarray}
\label{fipi2}
\partial_\varphi F_1= \partial_\varphi F_2=  \partial_\varphi F_3= \partial_\varphi F_0=0,~
\partial_\varphi S_1=   S_2=   S_3=0.
\end{eqnarray}

In order to obtain simple boundary conditions at the event horizon, 
 it proves useful to introduce a new (compact) radial
coordinate  $x$, with\footnote{Numerics are performed by employing the compactified coordinate $x$.}
\begin{eqnarray}
\label{x}
r=\frac{r_H}{1-x^2}
\end{eqnarray}
with range $0\leqslant x<1$, 
such that the horizon is located at $x=0$.
This results in the following boundary conditions at the horizon: 
\begin{eqnarray}
\label{eh}
\partial_x F_1=\partial_x F_2=\partial_x F_3=\partial_x F_0=0,~
S_1=S_2=\partial_x S_3=0.
\end{eqnarray}

 \subsection{Horizon quantities}

The induced  metric on the spatial sections of the horizon is
\begin{eqnarray}
\label{horizon-geom}
d\sigma^2=r_H^2 \left( 
F_2(r_H,\theta,\varphi)d\theta^2+  F_3(r_H,\theta,\varphi)(\sin \theta d\varphi+  S_3(r_H,\theta,\varphi)d\theta)^2
\right).
\end{eqnarray}
The horizon area and the Hawking temperature  are
\begin{eqnarray}
A_H=r_H^2 \int_0^{\pi} d\theta \int_0^{2\pi} \sin\theta \sqrt{F_2(r_H,\theta,\varphi)F_3(r_H,\theta,\varphi)}, \qquad 
T_H= \frac{1}{4\pi r_H}\left(1+\frac{3r_H^2}{L^2}-q^2 \right).
\end{eqnarray}
The  horizon has a spherical topology, as one can see by evaluating its Euler characteristic\footnote{This provides a further 
test of the 
accuracy of solutions,
as the (numerical) integral of the horizon Ricci scalar over the
horizon should equal $\chi$.} $\chi$. 
However, the deviation from sphericity can be significant, in particular 
for the EM-AdS BHs in the next Section.
A measure of this deviation can be found $e.g.$
by evaluating  
the circumference of the horizon along the equator
\begin{eqnarray}
L_e= r_H\int_0^{2\pi} d\varphi \sqrt{F_3(r_H,\pi/2,\varphi) },
\end{eqnarray}
together with 
the circumference of the horizon along the poles (which is $\varphi$-dependent):
\begin{eqnarray}
L_p (\varphi)=2 r_H\int_0^{ \pi} d\theta \sqrt{F_2(r_H,\theta,\varphi)+F_3(r_H,\theta,\varphi)S_3^2(r_H,\theta,\varphi) }.
\end{eqnarray}
 Also,
in principle,
the horizon geometry  (\ref{horizon-geom})
can be visualised
by considering its isometric embedding
in a three-dimensional Euclidean space, with
\begin{eqnarray}
 ds_E^2  =
dX^2 + dY^2 + dZ^2.
\end{eqnarray}
The embedding functions
$X(r,\theta)$,
$Y(r,\theta)$,
$Z(r,\theta)$
are found by integrating the following system of
non-linear PDEs 
\begin{eqnarray}
\nonumber
&&
X_{,\theta}^2+Y_{,\theta}^2+Z_{,\theta}^2=r_H^2 \left(
 F_2(r_H,\theta,\varphi)+F_3(r_H,\theta,\varphi) S_3^2(r_H,\theta,\varphi)
\right)
\\
&&
X_{,\varphi}^2+Y_{,\varphi}^2+Z_{,\varphi}^2=r_H^2 F_3(r_H,\theta,\varphi)\sin^2 \theta,
\\
\nonumber
&&
X_{,\theta}X_{,\varphi} +Y_{,\theta} Y_{,\varphi} +Z_{,\theta} Z_{,\varphi}  
=r_H^2 \sin \theta  F_3(r_H,\theta,\varphi)S_3(r_H,\theta,\varphi).
\end{eqnarray}
One remarks, however, that in general, these global isometric embeddings could only
be obtained up to some threshold configuration, beyond
which an obstruction, similar to the one found for fast spinning Kerr BH
\cite{Smarr:1973zz,Gibbons:2009qe},
 arises.

\subsection{The numerical approach} 
 
The EDT equations (\ref{EDT}) together with the matter field(s) equations can be solved by using various approaches. 
For all cases discussed in this work, the numerical  methods we have chosen to use
can be summarized as follows.
First, the equations are  discretized on a $(r, \theta,\varphi)$-grid with 
$N_r\times N_\theta \times N_\varphi$
points\footnote{
Typical grids have sizes around
$100 \times 30 \times 30$ points.
The grid spacing in the $r$-direction is non-uniform; but the angular grids are uniform.
}
The resulting system is then solved
iteratively until convergence is achieved. 
All numerical calculations for axially symmetric configurations
are performed by using a  professional software based on the iterative Newton-Raphson
method 
\cite{schoen}.
 This code requests the system of non-linear
PDEs to be written in the form 
$F(r, \theta,\varphi; 
u; u_r, u_\theta,u_\varphi; 
u_{rr}, u_{\theta\theta},u_{\varphi\varphi},
u_{r\theta},u_{r\varphi},u_{\theta\varphi}) = 0$, 
(where $u$ denotes
the set of unknown functions) subject to a set of boundary conditions on a rectangular domain. 
The user must
deliver the equations, the boundary conditions, and the Jacobian matrices for the equations
and the boundary conditions. 
Starting with a guess solution, small corrections are computed until a desired
accuracy is reached. The code automatically provides also an error estimate for each unknown function,
which is the maximum of the discretization error divided by the maximum of the function.
For most of the solutions reported in this work,
 the typical numerical error for the functions is estimated to be lower than
$10^{-3}$.

\section{The first mechanism: Einstein--Maxwell--AdS BHs}
\label{sec_1}
 
\subsection{Maxwell `solitons' in AdS spacetime}
   
For a Minkowski spacetime background, the existence of solitons in a given model
is supported by 
non-linearities of the field(s).
However, this is not necessarily the case for different spacetime asymptotics.
Perhaps the simplest example in this direction is provided by a Maxwell field in 
a globally AdS fixed geometry, with the usual action
\begin{eqnarray}
I_A = -\frac{1}{4}\int d^4 x\sqrt{-g}
F_{\mu \nu}F^{\mu\nu}
,
\end{eqnarray}
where
$
F_{\mu \nu}=\partial_\mu A_\nu-\partial_\nu A_\mu
$
is the U(1) field strength.
The 4-potential $A_\mu$ satisfies the Maxwell equations
\begin{eqnarray}
\nabla_\mu F^{\mu \nu}=0,
\end{eqnarray}
with an energy-momentum tensor 
\begin{eqnarray}
\label{TMik}
T_{\mu\nu}=F_{\mu \alpha}F_{\nu\beta}g^{\alpha \beta}-\frac{1}{4}g_{\mu\nu}F^2,
\end{eqnarray}
where $\rho=-T_t^t$ is taken as the energy density.

For the geometry, we
 consider a static spherically symmetric background, with a line element
\begin{eqnarray}
\label{AdS}
ds^2=-N(r)dt^2+\frac{dr^2}{N(r)}+r^2(d \theta^2+\sin^2\theta d\varphi^2)
, 
\end{eqnarray}  
where $r,\theta,\varphi$
are spherical coordinates.
Globally AdS spacetime corresponds to  taking
\begin{eqnarray}
\label{N-AdS}
N(r)=1 +\frac{r^2}{L^2},
\end{eqnarray} 
in  (\ref{AdS}), where $L$ is the AdS length scale, the cosmological constant
being $\Lambda=-3/L^2$.

 Given a static solution of the Maxwell equations, one can
define 
a total mass-energy $E$ and an electric charge $Q_e$ as
\begin{eqnarray}
E=-\int d^3 x \sqrt{-g}T_t^t,~~Q_e=\oint_{\infty}dS_r F^{rt}~.
\end{eqnarray}

Restricting to a purely electric U(1) potential   with
\begin{eqnarray}
\label{gauge-ansatz}
A= V(r,\theta,\varphi)dt,
\end{eqnarray} 
 and
assuming separation of variables for the electric potential,
a generic configuration has the expression
\begin{eqnarray}
\label{deco}
V(r,\theta,\varphi)= \sum_{\ell \geqslant 0}\sum_{m=-\ell}^{m=\ell}c_{\ell m}R_\ell(r)Y_{\ell m}(\theta,\varphi),
\end{eqnarray}
where $c_{\ell m}$ are
arbitrary constant and $Y_{\ell m}(\theta,\varphi)$ are the $real$ spherical harmonics.
%
For each $\ell,m$,
the radial function $R_\ell$ is a solution of the equation
\begin{eqnarray}
\label{eq}
 (r^2R_\ell')'=\frac{1}{N(r)}\ell(\ell+1)R_\ell,
\end{eqnarray}
where a prime denotes the derivative $w.r.t.$ the radial coordinate.
Note that the integer $m$ does not enter the above equation. 

For a Minkowski spacetime background $N(r)=1$, the general solution of the equation (\ref{eq}) reads
\begin{eqnarray}
\label{solM}
 R_\ell(r)=c_1 r^\ell+\frac{c_2}{r^{\ell+1}},
\end{eqnarray}
with $c_{1,2}$ arbitrary constants.
  Thus, any non-trivial solution diverges either at the origin or at infinity;
also, the total energy associated to any such multipolar field is infinite. 
However,  
``boxing" Minkowski spacetime allows for everywhere regular electric multipoles,
 with a finite total energy inside the box. 
That is, 
physically reasonable configurations can be obtained by taking $c_1=0$ in (\ref{solM}),
and confining the electric field to be inside a box -- say, spherical and of radius $r_B$.  
 
In gravitational physics, AdS is a natural ``box", in view of its conformal timelike boundary. 
This suggest the existence of solutions of Maxwell equations in a (globally) AdS background 
which are  everywhere regular,
 with a finite total energy.
Indeed, this conjecture has been confirmed in 
\cite{Herdeiro:2015vaa}.
For any $\ell\geqslant 1$ the equation (\ref{eq})   possesses a solution\footnote{The solution is normalized such that $R_\ell (r)\to 1$ asymptotically.} 
which is regular everywhere, in particular at $r=0$:
\begin{eqnarray}
\label{solMgen}
R_\ell(r)=
\frac{\Gamma(\frac{\ell+l}{2})\Gamma(\frac{3+\ell}{2})}{\sqrt{\pi}\Gamma(\frac{3}{2}+\ell)}
 \frac{r^{\ell}}{L^\ell}~{}_2F_1\left(\frac{1+\ell}{2}, \frac{ \ell}{2},\frac{3}{2}+\ell,- \frac{r^2}{L^2}\right),
\end{eqnarray} 
where ${}_2F_1$ is the hypergeomtric function.
The simplified forms of the first two such functions are
\begin{eqnarray}
\nonumber
R_1(r)=-\frac{2}{\pi}\left[\frac{L}{r}~\left(-1+\frac{L}{r}\right)\arctan\left(\frac{r}{L}\right)\right], \qquad 
~~
R_2(r)=1+\frac{3L^2}{2r^2}-\frac{3L(r^2+L^2)}{2r^3}\arctan\left(\frac{r}{L}\right).
\end{eqnarray} 
 
 The asymptotic form of the general solution reads
 \begin{eqnarray}
R_\ell(r)=c_0 \left(\frac{r}{L}\right)^\ell+\dots ~~{\rm as}~~r\to 0, \qquad ~~{\rm and} \qquad ~~R_\ell (r)=1-c_1 \frac{L}{r}+\dots~~{\rm as}~~r\to \infty,
\end{eqnarray} 
where $c_0$, $c_1$ are $\ell-$dependent constants~\cite{Herdeiro:2015vaa}.

One can easily see that the energy density  $\rho$ 
 of the solutions is finite everywhere,
being strongly localized in a finite region of space,
and
depending on $both$ $\theta,\varphi$ in a complicated way.
In particular, for $m\neq 0$, a surface of constant energy density
possesses discrete symmetries only  \cite{Herdeiro:2016plq}.
Also,  $\rho$ is nonzero at $\theta=0$;
and
  it vanishes at the origin unless $\ell=1$.
At infinity, $\rho$ decays as $1/r^4$, such that
 the   total energy of these solutions is finite 
 \begin{eqnarray}
E_\ell=  
L \frac{ \Gamma(\frac{1+\ell}{2}) \Gamma(\frac{3+\ell}{2})} {\Gamma(1+\frac{\ell}{2}) \Gamma(\frac{\ell}{2})}~.
\end{eqnarray} 

By using a simple scaling argument, one can show that these static   Maxwell 
configurations in a fixed AdS
spacetime satisfy the virial identity
   \begin{eqnarray}
 \int_0^\infty r^2 dr \int_0^\pi \sin \theta
 \left (
 V_{,r}^2+\frac{1-\frac{r^2}{L^2}}{N^2(r)r^2}\left( V_{,\theta}^2+\frac{1}{\sin^2\theta} V_{,\varphi}^2\right)
\right )=0.
\end{eqnarray}
 This makes it clear that the AdS geometry supplies the attractive force needed to balance the repulsive force
of the gauge interactions.
Also, it is clear that the configurations are supported by the nontrivial angular dependence of $V$,
$i.e.$ they should possess a multipolar structure.

Finally, let us also mention that,
due to the electric-magnetic duality of Maxwell's theory in four dimensions, 
these configurations   possess also an equivalent purely magnetic picture 
\cite{Herdeiro:2016xnp}.

\subsection{Adding a BH horizon. Electrostatics in Schwarzschild-AdS background }
 
A rather similar picture is found when taking instead a Schwarzschild-AdS (SAdS) BH backgound, 
with a line element still given by 
(\ref{AdS}), where this time
\begin{eqnarray}
N(r)=
\left( 1-\frac{r_H}{r} \right)
\left(1+\frac{r^2}{L^2}\left(1+  \frac{r_H}{r}+\frac{r_H^2}{r^2}\right) \right).
\end{eqnarray}

\begin{figure}[h!]
\begin{center}
\includegraphics[height=.26\textheight, angle =0]{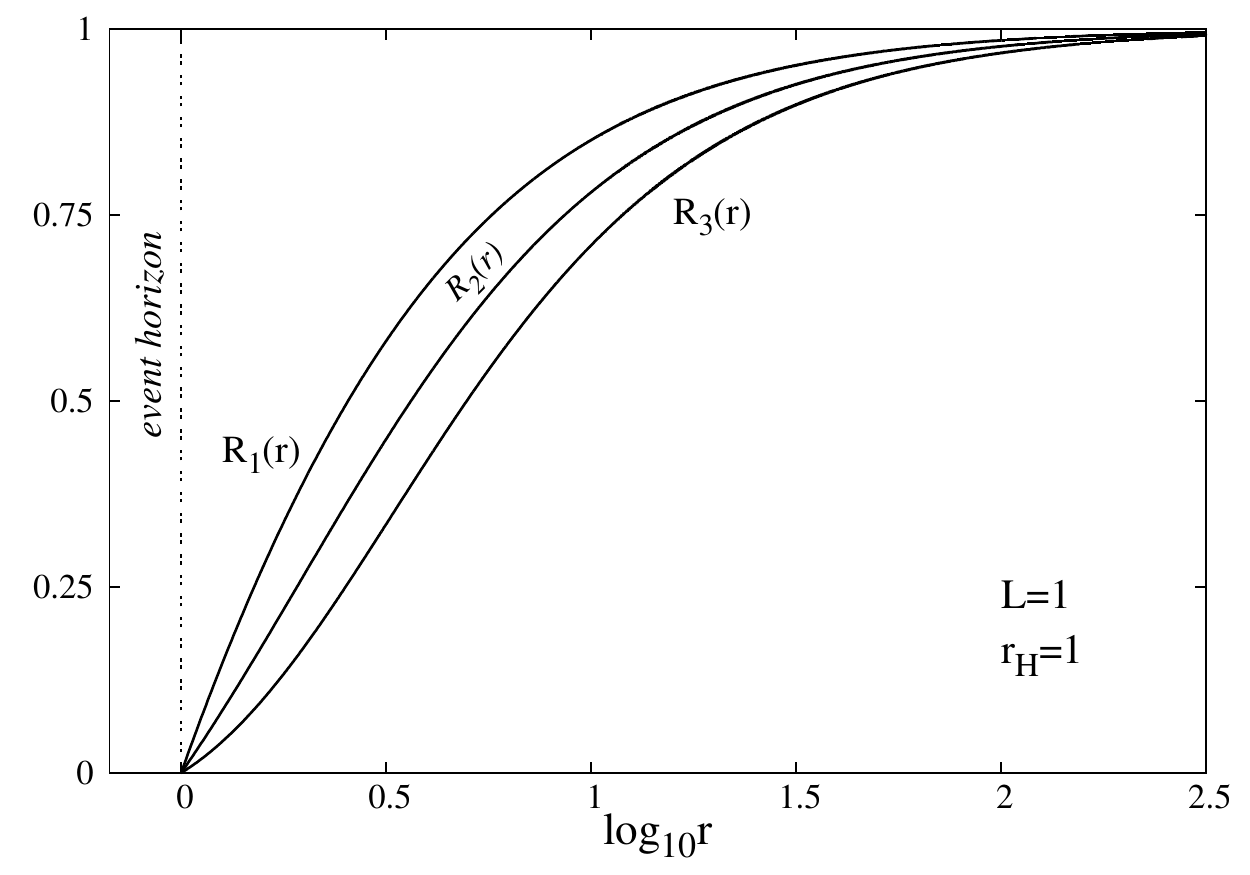} 
\includegraphics[height=.26\textheight, angle =0]{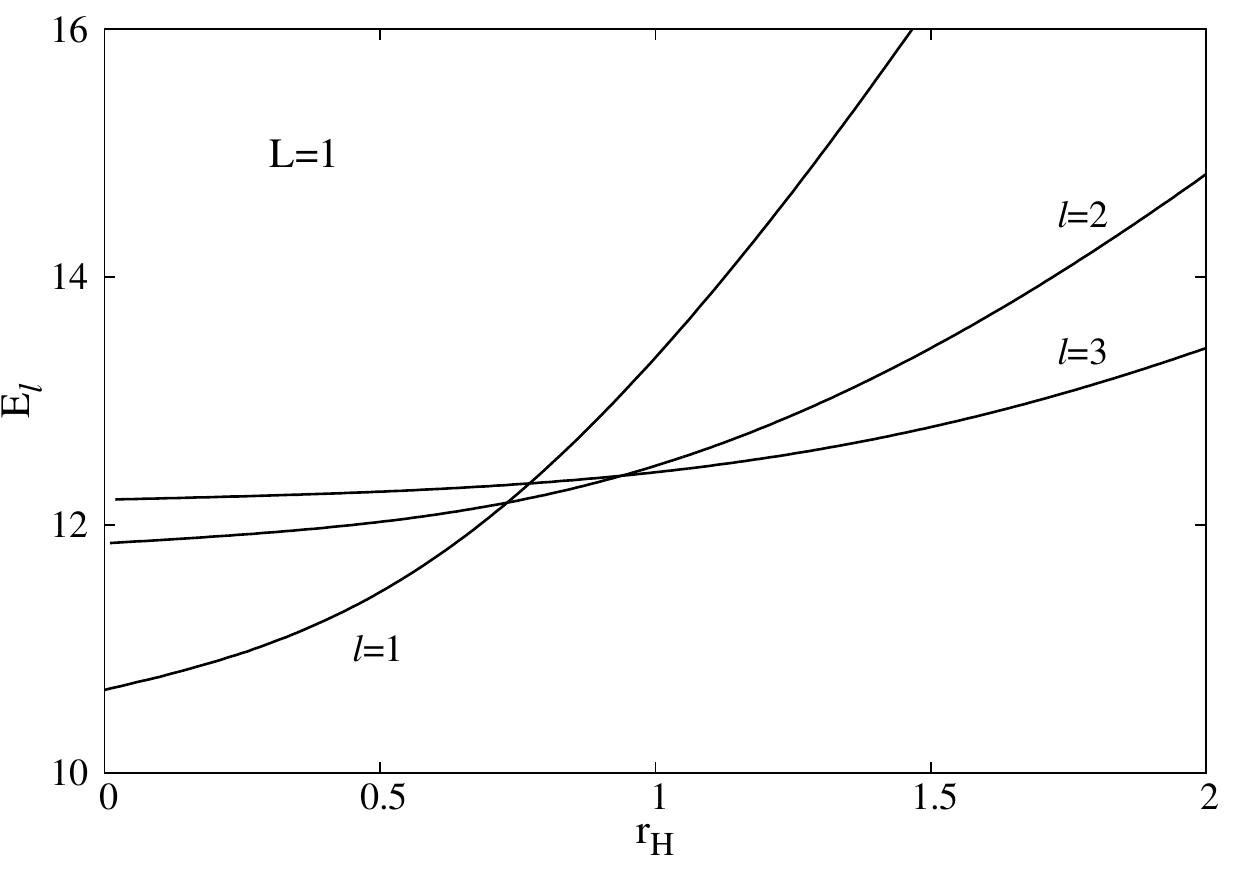} \ \ 
\end{center}
  \vspace{-0.5cm}
\caption{{\it Left panel:}  The radial function $R_\ell$ (with $\ell=1,2,3$)
 is shown  for Maxwell field solutions on a fixed Schwarzschild-AdS background.
{\it Right panel:} The  total mass-energy for families of $\ell=1,2,3,~m=0$ solutions
is shown as a function of the event horizon radius.  }
\label{Maxwell1}
\end{figure}
 
Unfortunately, equation (\ref{eq})  cannot be solved in closed form
in this case\footnote{However, an
exact solution can be found
for a Schwarzschild BH background, ($i.e.$ in the limit $L\to \infty$),
$R_\ell(r)$ 
 being the sum of two modes, one of them diverging at the horizon and the other one at infinity.
Thus, again, the ``boxing" feature of AdS spacetime regularizes the far field asymptotics.
},
except for  
  $\ell=0$, in which case $R_0(r)=c_0-{c_1}/{r}$. 
	However, an approximate expression of the solution can be found both near the horizon and in the far field. 
Assuming the existence of a  power series in $(r-r_H)$, one can easily see
from the eq. (\ref{eq}) 
 that, for 
 $\ell>0$,
the radial function 
necessarily
vanishes on the horizon\footnote{This results from the presence of the $N(r)$-factor in the denominator of the $r.h.s.$ in eq.~(\ref{eq}). Also, note that the condition $R_\ell(r_H)=0$, together with the expansion (\ref{asrh}), 
implies that the components of the  energy-momentum tensor (\ref{TMik})
are finite at the horizon.
}, the first terms in the solution therein 
  being 
\begin{eqnarray}
\label{asrh}
R_{\ell}(r)=r_1(r-r_H)+\frac{r_1\left ((\ell-1)(\ell+2)-\frac{6r_H^2}{L^2} \right)}{2r_H(1+\frac{3r_H^2}{L^2})}(r-r_H)^2+\mathcal{O}(r-r_H)^3,
\end{eqnarray}  
where $r_1$ is a parameter which results from the numerics.
As $r\to \infty$, the solution reads
\begin{eqnarray}
\label{asinf}
R_{\ell}(r)=1-c_1\frac{L}{r}+\frac{1}{2}\ell(\ell+1)\frac{L^2}{r^2}+\dots,
\end{eqnarray}
where we normalized it such that $R_{\ell}(r)\to 1$ asymptotically.

The $\ell\geqslant 1$ solutions  interpolating smoothly between the asymptotics (\ref{asrh}), (\ref{asinf})
are constructed numerically. 
In Figure~\ref{Maxwell1} (left panel) 
we exhibit the radial function $R_{\ell}$
for a SAdS background with a fixed horizon radius $r_H=1$ and $\ell=1,2,3$.
The dependence of the total mass-energy of the $\ell=1,2,3$; $m=0$ solutions as a function 
of the event horizon radius is shown in Figure \ref{Maxwell1} (right panel).
Note that in both plots we take an AdS length scale $L=1$.

\subsection{Including the back-reaction: Einstein-Maxwell-AdS BHs }
The existence of these everywhere regular, finite energy Maxwell fields, as described  above,
suggests 
that fully non-linear  Einstein-Maxwell-$AdS$ solitons, as well as deformed BHs, exist,
as the backreacting non-linear version  of the test field solutions.
In the absence of analytic methods\footnote{Closed form perturbative solitons  can be found,
as a power series in the parameter $c_e$ - see \cite{Herdeiro:2015vaa,Costa:2015gol,Herdeiro:2016xnp} for work in the axially symmetric case.
However, the  lowest order solutions are already extremely complicated.
 } to tackle the fully non-linear Einstein-Maxwell-AdS
solutions,
the problem is approached by using numerical methods,
as  described in Section  \ref{sec_0}.
Thus the EDT equations (\ref{EDT})
 are solved together with the Maxwell equations 
by employing
an electric
U(1) ansatz with
\begin{eqnarray}
\label{V}
A=V(r,\theta,\varphi) dt.
\end{eqnarray}
\begin{figure}[h!]
\begin{center}
\includegraphics[height=.27\textheight, angle =0]{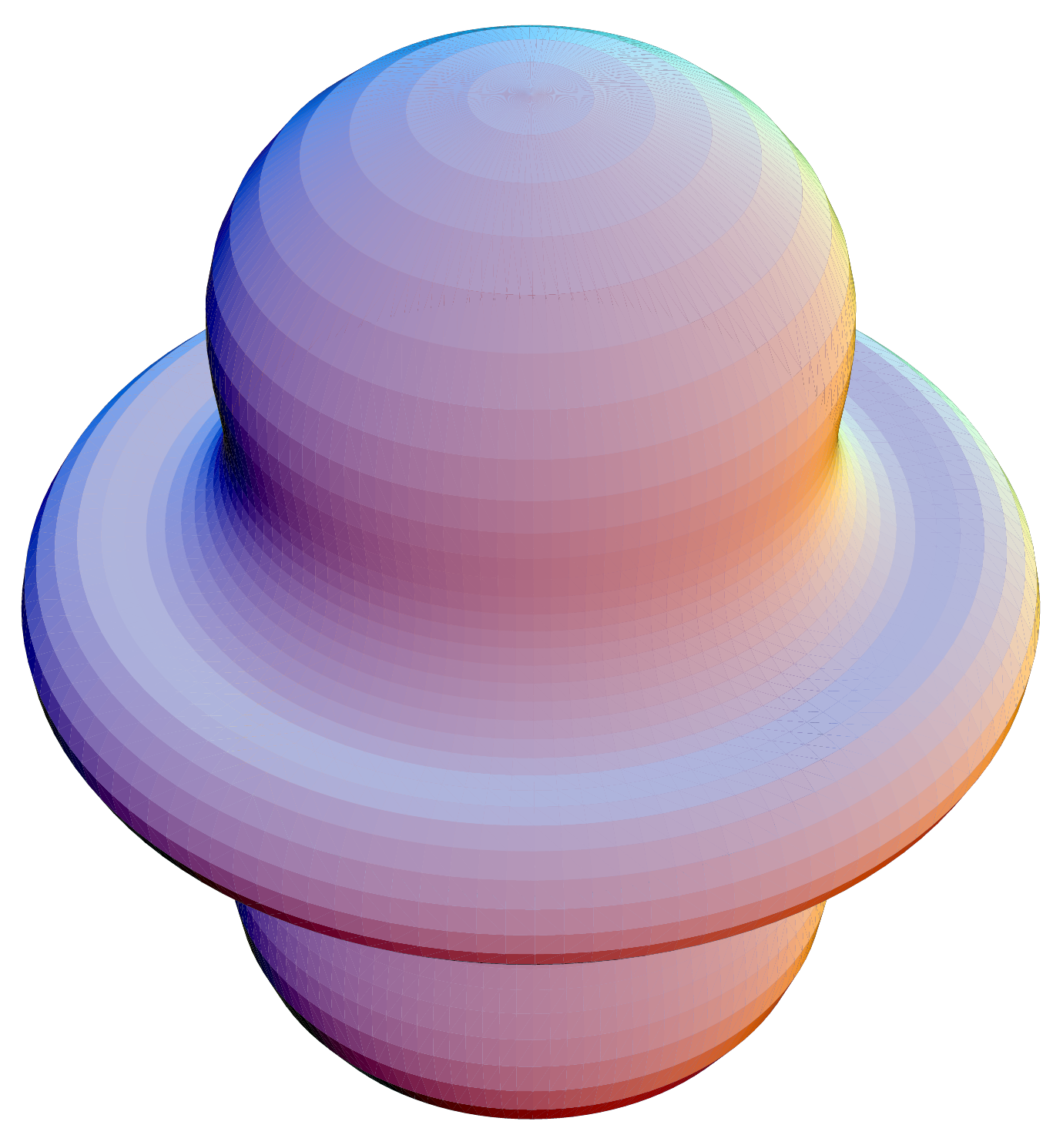} 
\includegraphics[height=.25\textheight, angle =0]{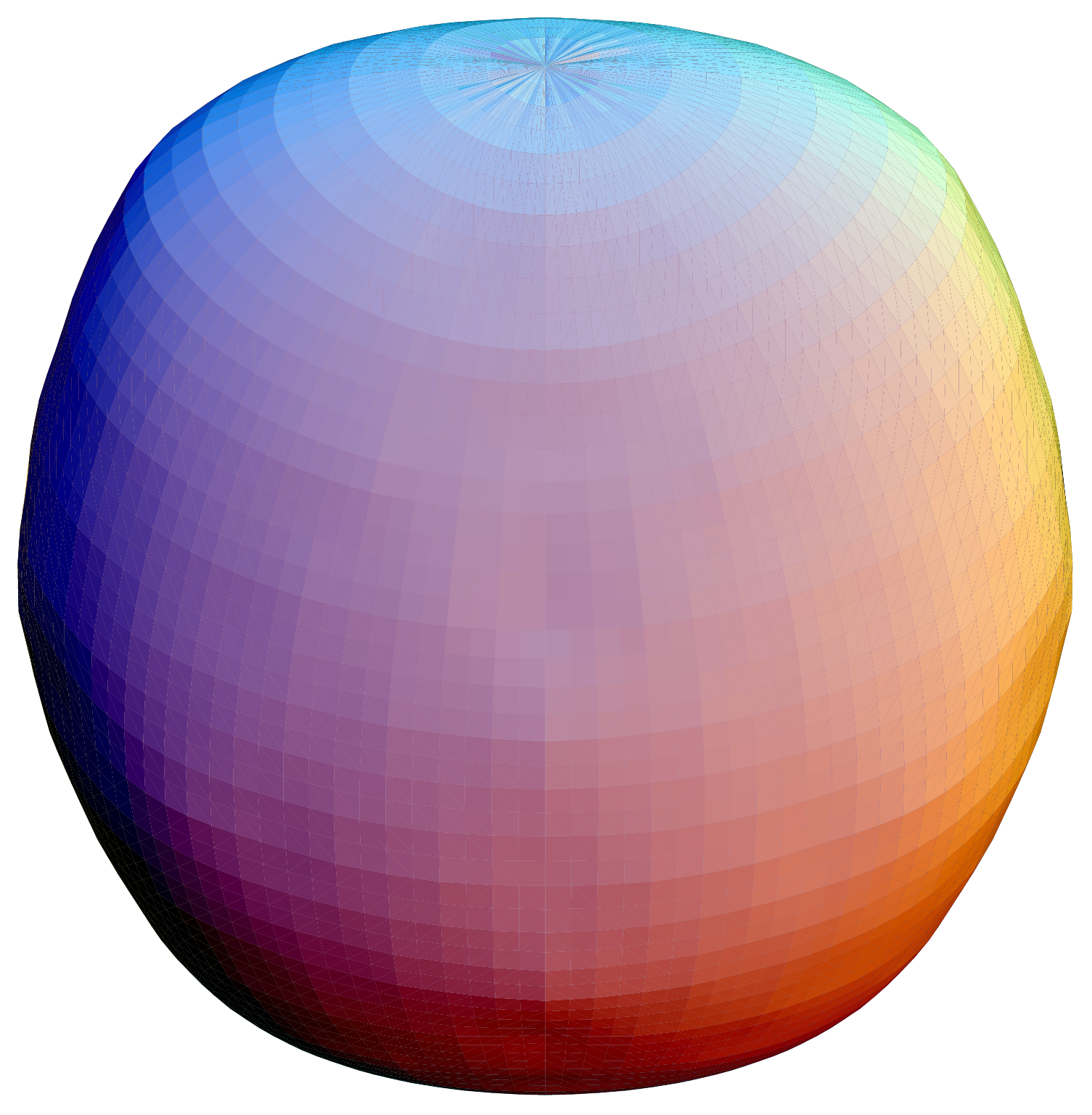} 
 \ \ 
\end{center}
  \vspace{-0.5cm}
\caption{ The isometric embeddings in Euclidean 3-space for the spatial sections of the horizon
of typical AdS-electrovacuum BHs with  ($\ell=2; m=0$) (left) and ($\ell=3; m=1$) (right).}
\label{horizon3d}
\end{figure}
%
\begin{figure}[h!]
\begin{center}
 \includegraphics[height=.26\textheight, angle =0]{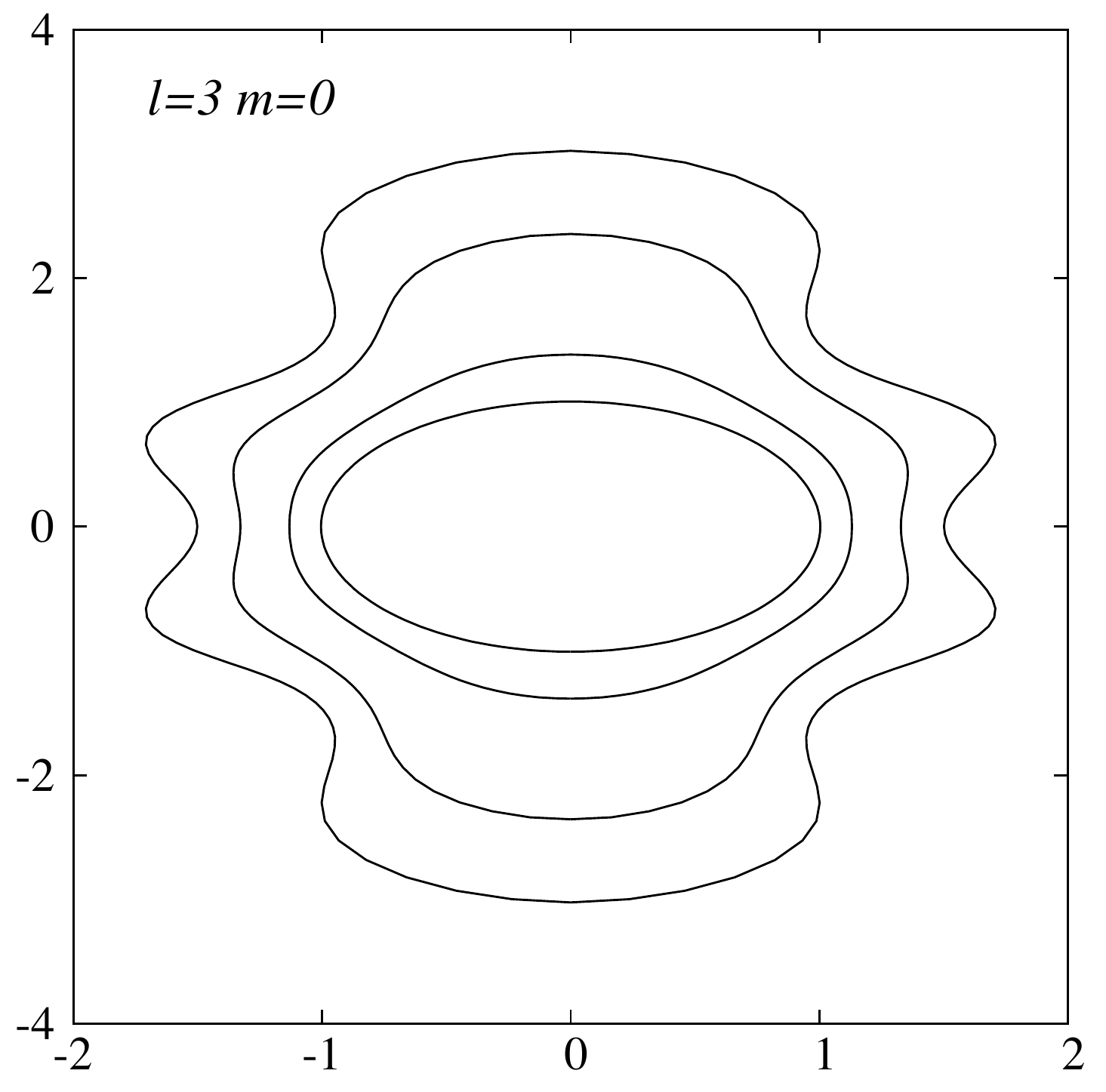}
 \includegraphics[height=.26\textheight, angle =0]{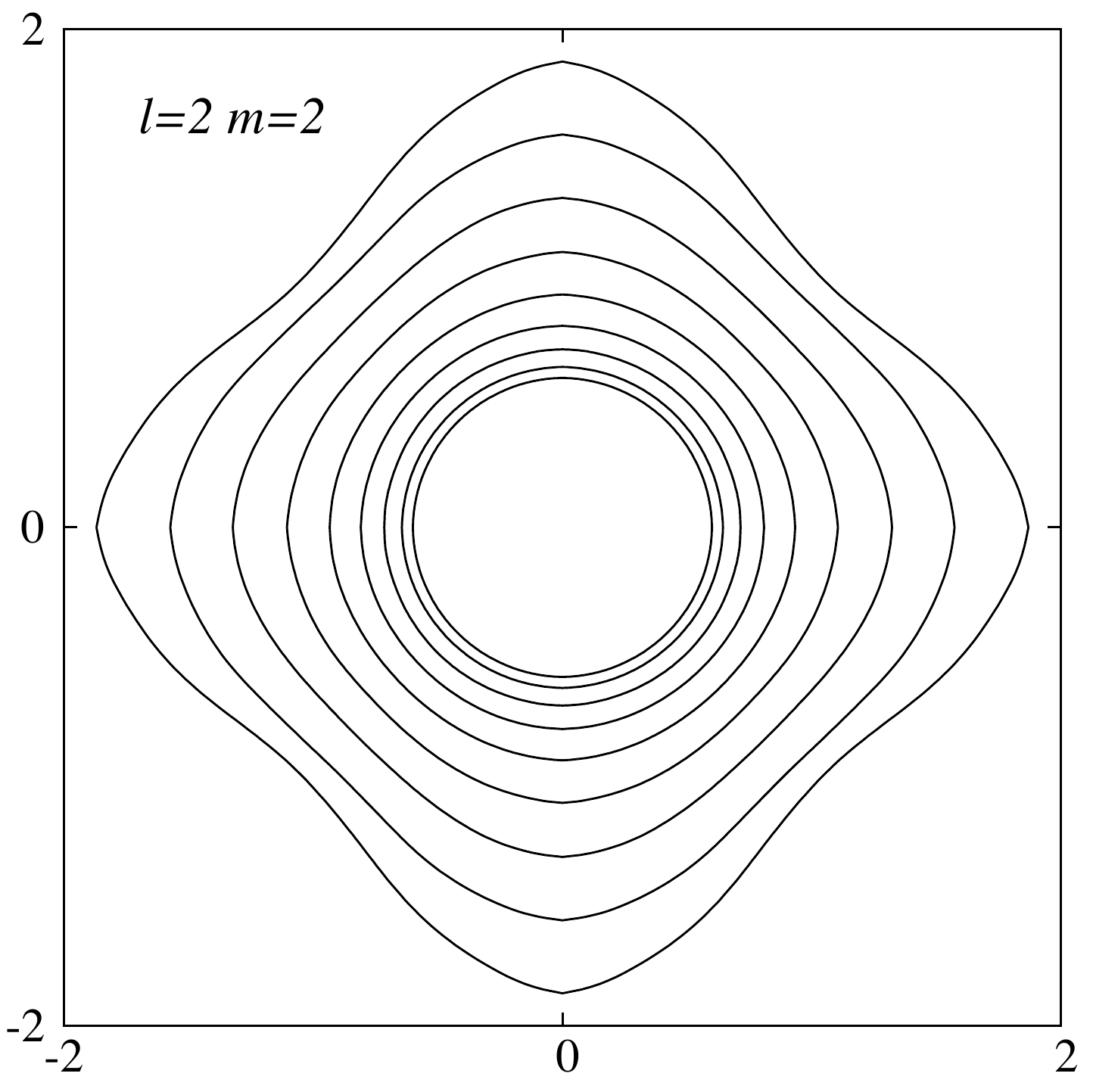} 
 \ \ 
\end{center}
  \vspace{-0.5cm}
\caption{  Equatorial slices for the horizon isometric embeddings of AdS-electrovacuum BHs with different boundary
data. The different lines correspond to BHs that have the same temperature, but increasing the
values of the parameter $c_e$. The right panel is extracted from~\cite{Herdeiro:2016plq}.}
\label{slice1}
\end{figure}

While the metric ansatz and the corresponding BCs are those displayed in Section 2,
the boundary conditions satisfied by the electrostatic potential are 
\begin{eqnarray}
\label{V-bc}
  V\big|_{r=0}=0,~~
	V\big|_{r\to \infty} =c_e Y_{\ell m}(\theta,\varphi),~~
V\big|_{\theta=0} =0,~~
   V\big|_{\theta=\pi/2}=0,~~
 \partial_\varphi  V\big|_{\varphi=0}=0,~~
  V\big|_{\varphi=\pi/2}=0.
\end{eqnarray}
with $c_e$ a constant.
Note that if $\ell+m$ an even number we impose instead 
$
\partial_\theta  V\big|_{\theta=\pi/2}=0
$;
also, for even $m$ we shall require
$
 \partial_\varphi  V\big|_{\varphi=\pi/2}=0,
$
as implied by the symmetries of the problem.
No upper bound on $c_e$ seems to exist, although the numerical accuracy decreases
for large values of this parameter.
For example, for the $(\ell=2,m=2)$ case,
the maximal considered value of $c_e$ was $15$,
while for  $(\ell=3,m=3)$  we have constructed solutions up to $c_e=12$.

The numerical solutions were reported in Ref. \cite{Herdeiro:2016plq};
here we review their basic properties.
First, the configurations which were found
in the probe limit for a SAdS geometry survive when including
backreaction on the spacetime geometry.
The case $(\ell=m=0)$
is special, corresponding to the RN-AdS BHs.
 Non-linear continuation exist for all $(\ell,m)$-modes.
While the solutions found starting with a $m=0$ mode 
are axially symmetric, static BHs without isometries
  exist as well, being the nonlinear continuation of the probe solutions with $m\neq 0$.
	A systematic study of the axially symmetric $(\ell=1,m=0)$
case was reported in Ref. \cite{Costa:2015gol},
	where the solutions were dubbed `{\it polarised BHs}',
	as justified by the local distribution of the electric charge.
	
These BHs possesses a single global charge, corresponding to their total mass, which is computed 
$e.g.$ by  employing either the prescriptions in Ref. \cite{Balasubramanian:1999re},
as described in Appendix A, while the net electric charge vanishes.
As such, most of the basic thermodynamical features 
of these configurations are similar to the (vacuum) 	SAdS case.
For example, the BH temperature   is always bounded from below.
At
low temperatures we have a single  solution, which  corresponds to the
thermal globally regular solution. 
At high temperatures there exist two additional solutions
that correspond to the small and large BHs \cite{Herdeiro:2016plq}. 
Also, 
the minimal temperature of the BHs decreases with the
 increase of
 $c_e$, although, at least for the explored solutions,
it never reaches zero.

Despite these SAdS-like features, the horizon geometry of these configurations
can strongly depart from spherical symmetry, as shown in Figures \ref{horizon3d}, \ref{slice1}.

\section{The second mechanism:  scalarized RN BHs}
\label{sec_2}

\subsection{The general setting and the zero mode}

The situation is rather different in this case. There is no solitonic limit 
whose basic properties (in particular the absence of isometries) are inherited by the 
BH generalizations~\cite{Herdeiro:2019oqp} (but see~\cite{Herdeiro:2019iwl}).
Instead,  
the mechanism at work here
has a different origin, residing on the phenomenon of
'spontaneous scalarization'
and
 the existence of $(\ell,m)$-{\it scalar clouds} for a given 
(electro-)vacuum configuration $S_0$.
 The non-linear continuation of these clouds results in a set  $S_e$ of static BHs without isometries.
These solutions violates the well-known no-hair theorems
due to the non-standard scalar field action.
	In a nutshell, the scalar field action to be considered can be expressed as 
\begin{eqnarray}
\label{actionS}
I_\phi= \int  d^4 x \sqrt{-g} 
\left[
 \frac{1}{2}(\nabla \phi)^2 
+f(\phi) {\cal I}(\psi;g)
\right],
\end{eqnarray}
with $f(\phi) $ some
{\it coupling function}
and 
${\cal I}$
a {\it source term} which generically depends on 
some extra-matter field(s)
$\psi$
and
metric tensor $g_{\mu\nu}$.
Then 
the corresponding equation of motion for the scalar field $\phi$
reads
\begin{eqnarray}
\label{eq-phi}
\nabla^2\phi=\frac{\partial f}{\partial \phi}{\cal I}.
\end{eqnarray}

Next, one
 assumes the existence of a ground state for the scalar field with
$ \phi=0$,
which is the fundamental solution of the equation (\ref{eq-phi}); then the coupling function should satisfy the condition\footnote{This condition excludes the case of a dilaton coupling, $f(\phi)=e^{-\alpha \phi}$.}
\begin{eqnarray}
\label{condx}
\frac{\partial f}{\partial \phi}\Big |_{\phi=0}=0.
\end{eqnarray}
Thus, the usual electrovacuum solutions solve the considered model,
(\ref{actionS}) supplemented with the Einstein-Hilbert action and, possibly, the action of other matter field(s) $\psi$. 
This provides  the {\it fundamental solutions} of the model.
Apart from that, the model possesses a second set of solutions, with a nontrivial scalar field -- {\it the scalarized BHs}.
These solutions are usually entropically preferred
over the fundamental ones
 ($i.e.$ they maximize the entropy for given global charges).
Moreover, they are smoothly connected with the fundamental set, approaching it for $\phi=0$.

At the linear level, spontaneous scalarization manifests
itself as a tachyonic instability triggered by a negative
effective mass squared of the scalar field.
This can be seen by considering the linearized form of the equation (\ref{eq-phi})
($i.e.$ with a small-$\phi$):
\begin{eqnarray}
\label{eq-phi-small}
(\nabla^2-\mu_{eff}^2)\phi =0,~~{\rm where}~~\frac{1}{2}\mu_{eff}^2=  \frac{\partial^2 f}{\partial \phi^2}\Big |_{\phi=0} {\cal I} .
\end{eqnarray}
Assuming 
a spherically symmetric background
as given by  (\ref{AdS}) 
and
a decomposition of the scalar field in spherical harmonics
similar to 
(\ref{deco}),
 equation (\ref{eq-phi-small})
implies
that the amplitude 
$R_{\ell}(r)$
is
a solution of the equation 
 \begin{eqnarray}
\label{eqf1}
\frac{1}{r^2} (r^2 N R_\ell')'= \left(\frac{\ell(\ell+1)}{r^2}+\mu_{eff}^2 \right)R_\ell.
\end{eqnarray} 
The solutions of the above equation describe scalar clouds.
In fact, solving (\ref{eqf1}) can be viewed as an eigenvalue problem:
for a given $\ell$,
the condition for  a smooth
scalar field which vanishes asymptotically selects a discrete set of $S_0$
 configurations.
 
\begin{figure}[h!]
\begin{center}
\includegraphics[width=0.5\textwidth]{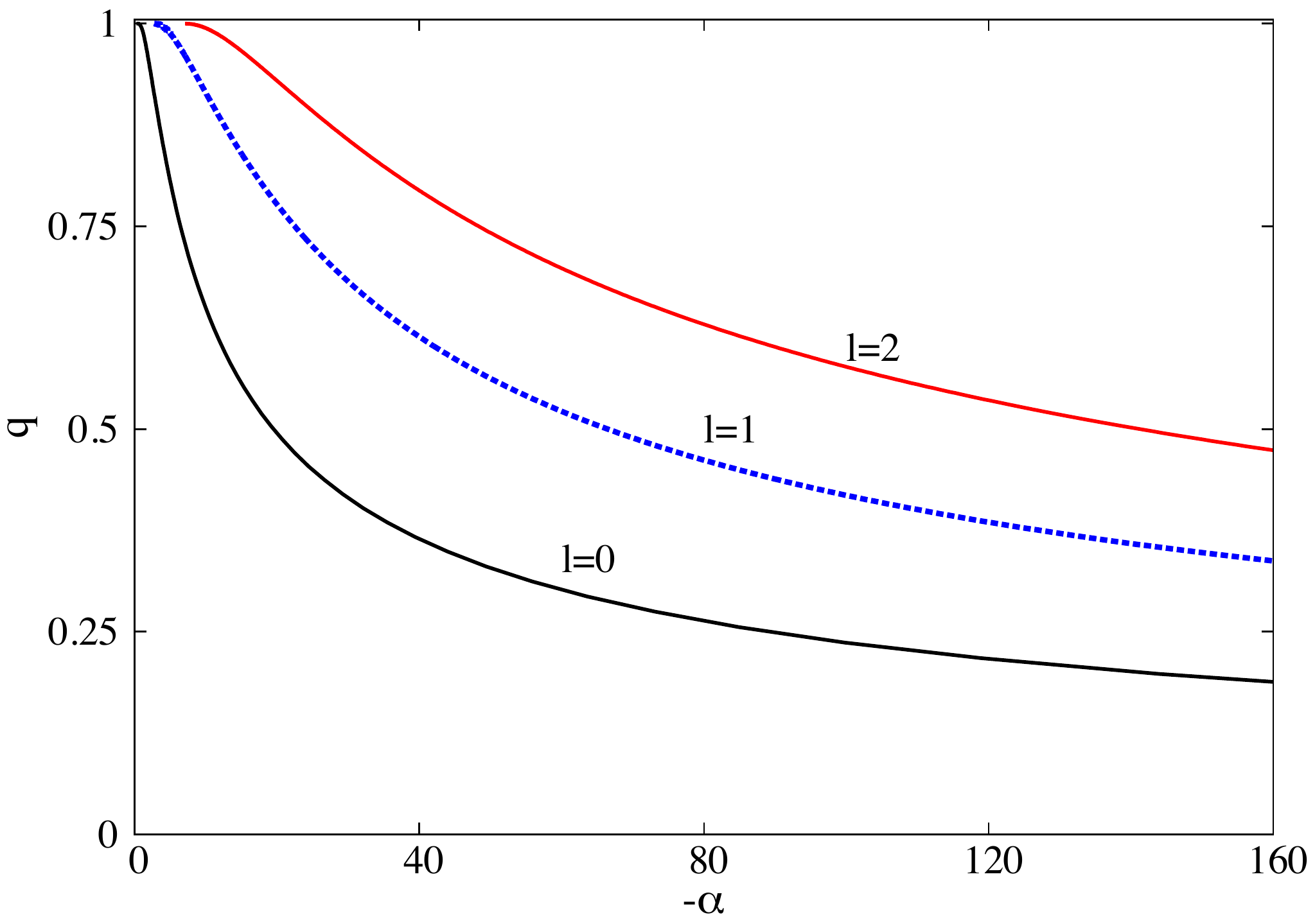}\ \ \
\caption{
Existence lines for scalar clouds on a RN background are shown as a function of the parameter $\alpha$,
for $\ell=0,1,2$.
For each $(\alpha,\ell)$, a branch of scalarised solutions bifurcates
from a RN BH with a particular charge to mass ratio $q=Q/M$.
}
\label{EMS0}
\end{center}
\end{figure}

In this setting, one can distinguish
 two types of scalarisation, 
depending on the  `source' term ${\cal I}$.
Geometric scalarisation started being considered in  Refs.
\cite{Silva:2017uqg,Doneva:2017bvd,Antoniou:2017acq},
using the Gauss-Bonnet invariant  
as the source term  
\begin{eqnarray}
\label{LGB}
 {\cal I}=L_{GB},
\end{eqnarray}
with $S_0$ the Schwarzschild BH.
Matter-induced scalarisation is illustrated by the work \cite{Herdeiro:2018wub}, 
which studied the
spontaneous scalarisation of electrovacuum BHs, and where the source term was
  \begin{eqnarray}
	\label{M}
 {\cal I}=F_{\mu \nu}F^{\mu \nu},
\end{eqnarray}
with $S_0$ the RN BH.
In both cases. 
the explicit form of the coupling function
does not appear for be important
as long as the condition (\ref{condx})
is satisfied. 
For concreteness, we shall present results for
 \begin{eqnarray}
 f(\phi)=e^{-\alpha \phi^2},
\end{eqnarray}
with the coupling constant $\alpha<0$ being an input parameter.

The main advantage of the matter-induced scalarisation 
model (\ref{M})
consists in its simplicity.
For example, 
  the equations are simple enough to allow a  non-perturbative study of the $general$ non-spherical solutions, which is not the case for model (\ref{LGB}).
Moreover, the zero-mode equation (\ref{eqf1}) has  $\mu_{eff}^2=\alpha Q^2/r^4$ 
and
can be solved  in closed form for $\ell=0$, with
\begin{eqnarray}
\label{ex1}
R _0(r)=P_u 
\left[
1+\frac{2Q^2(r-r_H)}{r(r_H^2-Q^2)}
\right],
\end{eqnarray}
where $u\equiv(\sqrt{4\alpha+1}-1)/2$, $r_H\equiv M+\sqrt{M^2-Q^2}$ and $P_u$ is a Legendre function
($(M,Q)$ being the mass and charge parameter of the RN BH).
Thus,
for generic
parameters  $(\alpha,Q,r_H)$, finding the $\ell=0$ bifurcation points from RN reduces to studying the zeros of this function 
as  $r\to \infty$, such that the condition $R _0 \to 0$ is satisfied.
Although the general $\ell\geqslant 1$ solution of the Eq. (\ref{eqf1})
is not known in closed form, the results in Figure \ref{EMS0}
indicated that the $\ell=0$
results are generic, although the minimal value of $|\alpha|$ increases with $\ell$. Some further analytic studies of these scalar clouds can be found in~\cite{Hod:2020ljo}.

\begin{figure}[h!]
\begin{center}
\includegraphics[width=0.5\textwidth]{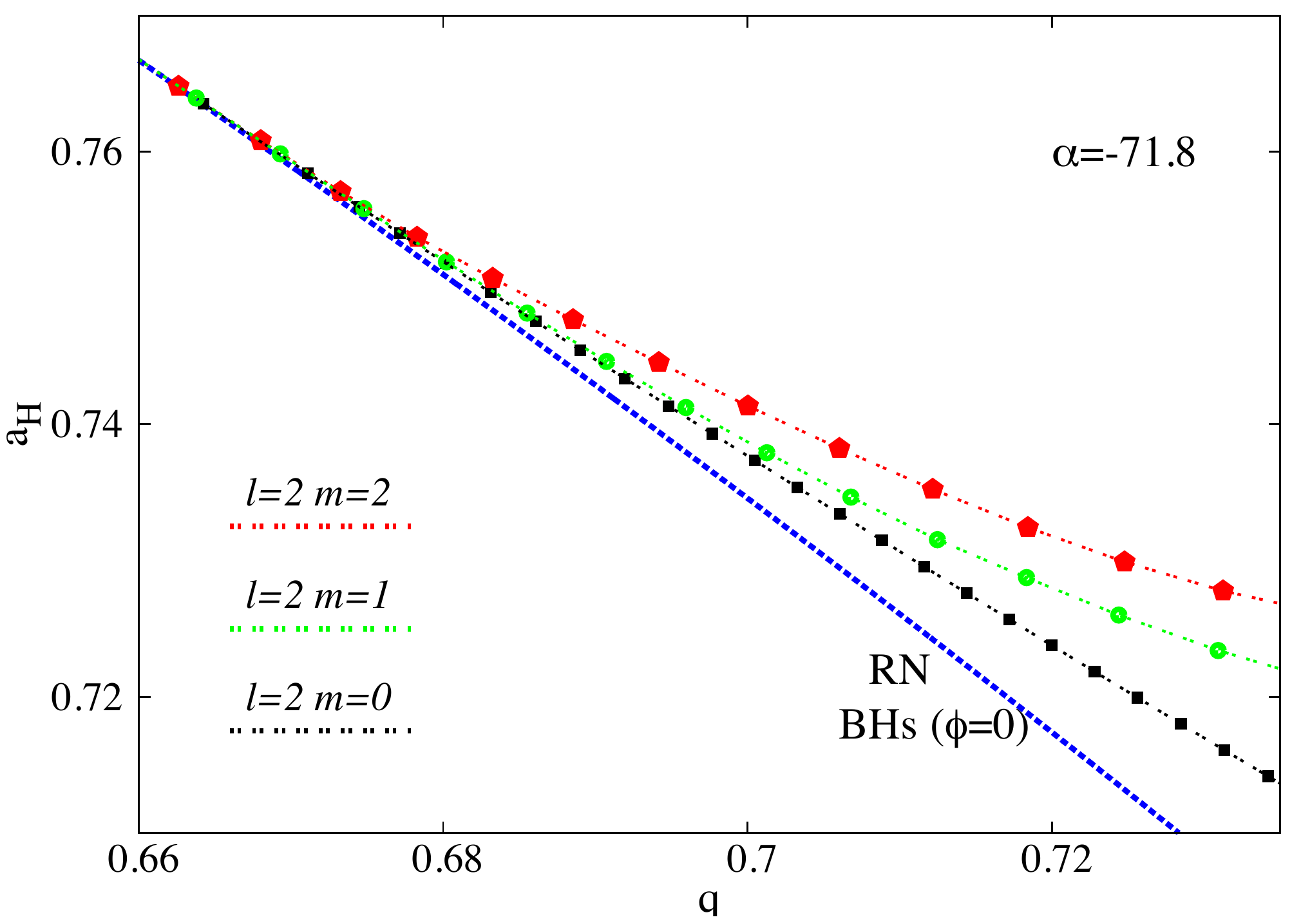}\ \ \
\caption{
Reduced area, $a_H\equiv \frac{A_H}{16\pi M^2}$, of scalarized BHs 
with $\alpha=-71.8$ $vs.$ reduced charge $q$. 
BHs with $m\neq 0$ have no spatial isometries. Extracted from~\cite{Herdeiro:2018wub}.
}
\label{EMS2}
\end{center}
\end{figure}

\subsection{Scalarized  BHs with no isometries}
The non-linear realisation of a general $(\ell,m)$-mode
results in a set $S_e$ of scalarized RN BHs.
These solution were reported in Ref. \cite{Herdeiro:2018wub} 
(see also Ref. \cite{Fernandes:2019rez} for a more detailed study of the spherical sector). 
They possess a variety of interesting properties, the most important one
being that some spherically symmetric scalarized BHs
emerge dynamically as end points of numerical evolutions
starting with small scalar perturbations of a RN BH.
Also, differently from the AdS BHs in the previous Section,
these configurations do not possess a  
solitonic limit.
 
In the context of this work,
of main interest are the configurations 
originating from 
 scalar clouds  with $m\neq 0$,
which
 result  in branches of static BH solutions with no isometries.
In particular, the horizon, despite being a topological two-sphere,
possesses discrete symmetries only.

Since no exact solution seems accessible (not even at the perturbative 
 level), these configurations are found numerically, by using the framework in Section 1.
The background metric is provided by the RN BH,  
with
 $N(r)$ given by the $L\to \infty$ limit in eq. (\ref{N}) (with $q=Q$). 
The ansatz for the Maxwell field is still given by (\ref{V}),
while for the scalar field the anstaz has the most general expression compatible with a static spacetime,
\begin{eqnarray}
\label{phi-gen}
\phi=\phi(r,\theta,\varphi).
\end{eqnarray}
In numerics, we impose the following BCs for the matter fields 
\begin{eqnarray} 
&&
\nonumber
~V\big|_{x=0}=0,~\partial_x \phi\big|_{x=0}=0,~~
 \partial_x V\big|_{x=1}=Q, ~\phi\big|_{x=1}=0,~~
\partial_\theta V \big|_{\theta=0}=\phi\big|_{\theta=0}=0,~~
\partial_\theta V \big|_{\theta=\pi/2}=\phi\big|_{\theta=\pi/2}=0,~~
\\
&&
\partial_\varphi  V\big|_{\varphi=0}=0 ,~~  \phi\big|_{\varphi=0}=0,~~
\partial_\varphi  V\big|_{\varphi=\pi/2}=0 ,~~  \phi\big|_{\varphi=\pi/2}=0,
\end{eqnarray} 
while for even $m$ we impose instead
$\partial_\varphi\phi\big|_{\varphi=0,\pi/2}  =0$; $Q$ is an input parameter fixing the electric charge,
and we recall that  $r=r_H/(1-x^2)$.

The only global charges are the  mass and the electric charge,
which 
are read off from the far field asymptotics
\begin{eqnarray}
\label{eqs1s}
-g_{tt}=F_0N=1-\frac{2M}{r}+\dots,~~V=\Phi-\frac{Q}{r}+\dots
\end{eqnarray}
with $\Phi$ the electrostatic potential.
A general large-$r$ expression of the solutions can
be constructed as a series in $1/r$, with $e.g.$ $F_0=1+c_t/r+\dots$, where $c_t$ is a constant.

 The main properties of the solutions can be summarized as follows.
First,
all configurations  are regular on and outside the horizon
and approach asymptotically 
the Minkowski spacetime background. 
Although the  deformation from sphericity is much less pronounced than in the 
AdS case discussed in the previous Section, 
the horizon geometry differs from that of the RN BH, possessing no isometries.
Perhaps
the most interesting result is that, for given $m$,
the entropy is maximized by the solutions emerging from the $\ell=m$ zero mode, see Figure \ref{EMS2}. 
However, in a full diagram, the entropy of the solutions with a given mass and electric charge
is maximized by the spherically symmetric ($l=m=0$) scalarized BHs \cite{Herdeiro:2018wub}.

\section{Further remarks}
\label{sec_3}
The main purpose of this work was to propose a general framework 
for the investigation of static BHs without isometries, together with
two different physical mechanisms allowing for 
such configurations. 
The first mechanism is based on the existence in some models  of solitons  without isometries,
while the second one relies on BH scalarization.
Explicit realizations were considered in both cases, corresponding to Einstein-Maxwell-AdS  
and Einstein-Maxwell-scalar models, respectively. 
These BHs have a
smooth, topologically spherical horizon, but without isometries, and approach, asymptotically,
the AdS or the Minkowski spacetime backgrounds.

\medskip
 
We expect similar solutions to exist in a variety of other models.
As such, the simple picture found in the electrovacuum case 
\cite{Israel:1967wq,Israel:1967za}, cannot be taken as a rule; in more general models, staticity does not guarantee the existence
of any continuous spatial symmetry, for physically
acceptable BH solutions.

For example, concerning the first mechanism,
we expect the known field theory solutions
without isometries in
Refs. \cite{Houghton:1995bs,Faddeev:1996zj,Battye:1998zn,Battye:1997qq} 
to possess BH generalizations.
In principle, these solutions can be constructed within the framework
proposed here, the only obvious obstacle  being the complexity of the (highly nonlinear) matter field
equations\footnote{For example,  the issue of gauge fixing for non-Abelian
fields with a nontrivial dependence of all space coordinates for a numerical approach is an open problem.
The same problem for the axially symmetric case is non-trivial, 
possessing a number of subtleties \cite{Brihaye:1992jk}. 
}. 
In this context, we 
mention the existence of asymptotically flat, static BHs without isometries 
in a simple model consisting in Einstein gravity coupled with a self-interacting scalar field,
a much simpler case than the matter field models in Refs. \cite{Houghton:1995bs,Faddeev:1996zj,Battye:1998zn,Battye:1997qq}.
Static and axially symmetric BHs solutions of this model
were constructed in Ref. \cite{Kleihaus:2013tba}.
They evade the no-hair theorems by having a scalar potential which is not strictly positive,
and possess a solitonic limit. We have found that the same mechanism allows for the existence of solitons and BHs without spatial isometries,
which will be discussed elsewhere.

\medskip
Concerning the second mechanism, we would like to comment on the results in the recent work 
\cite{Collodel:2019kkx},
which deals with BH scalarization in Einstein-Gauss-Bonnet-scalar theory,
$i.e.$ a source term (\ref{LGB}).
The results there indicate that, at least for the $(\ell\neq 0,~m=0)$
case, 
some basic features found for scalarized RN BHs are generic,
with the existence of static, axially symmetric solutions.
These configurations can be viewed as non-linear continuations of the corresponding zero-mode solutions
of the equation (\ref{eqf1}).
As remarked in \cite{Collodel:2019kkx},
more general static configurations without rotational symmetry should also exist,
the only obstacle in their study being the tremendous complexity of the gravity equations
in the presence of a Gauss-Bonnet term.

Still on the same subject, we mention the existence in some models of a similar instability
of Schwarzschild/RN BHs with respect to higher spin fields.
An  interesting case here are the pure Einstein-Weyl gravity solutions in \cite{Lu:2015cqa}.
The configurations there are spherically symmetric, bifurcating from a critical Schwarzschild BH, $i.e.$ with an $(\ell=m=0)$
zero mode.
Similar configurations are likely to exist for the higher 
$(\ell,m)$ case, which would correspond
 to static BHs without isometries in a pure gravity model.


\section*{Acknowlegements}
This  work  is  supported  by  the Center  for  Research  and  Development  in  Mathematics  and  Applications  (CIDMA)  
through  the Portuguese Foundation for Science and Technology (FCT - Fundacao para a Ci\^encia e a Tecnologia), references UIDB/04106/2020 and UIDP/04106/2020 and by national funds (OE), through FCT, I.P., in the scope of the framework contract foreseen in the numbers 4, 5 and 6 of the article 23, of the Decree-Law 57/2016, of August 29, changed by Law 57/2017, of July 19.  We acknowledge support  from  the  projects  PTDC/FIS-OUT/28407/2017  and  CERN/FIS-PAR/0027/2019.   This work has further been supported by the European Union Horizon 2020 research and innovation (RISE) programme H2020-MSCA-RISE-2017 Grant No. FunFiCO-777740.  
The authors would like to acknowledge networking support by the COST Action CA16104.

\bigskip

 \appendix
\section{Einstein-Maxwell-AdS BHs: far field asymptotics and boundary stress tensor}

The general Einstein-Maxwell-AdS solutions discussed
in Section \ref{sec_1} possess a
far field expansion 
with the  following leading order terms
(with $\alpha^2=4\pi G$):
\begin{eqnarray}
\nonumber
&&
V(r,\theta,\varphi)=v_0(\theta,\varphi)+v_1(\theta,\varphi)\frac{L}{r}-\frac{1}{2}
\left(v_{0,\theta\theta}(\theta,\varphi)+\cot\theta v_{0,\theta}(\theta,\varphi)
      +\frac{1}{\sin^2 \theta }v_{0,\varphi\varphi}(\theta,\varphi)\right)
\left(\frac{L}{r}\right)^2+\dots,
\\
\nonumber
&&
F_1(r,\theta,\varphi)=1+\left(\frac{q^2}{L^2}+\alpha^2 \left(v_{0,\theta}^2(\theta,\varphi)+\frac{1}{\sin^2\theta}v_{0,\varphi}(\theta,\varphi)^2\right) \right)
\left(\frac{L}{r}\right)^4+\dots
\\
\label{asympt1}
&&
F_2(r,\theta,\varphi)=1+f_{23}(\theta,\varphi)\left(\frac{L}{r}\right)^3-\alpha^2 \frac{1}{\sin^2\theta}v_{0,\varphi}(\theta,\varphi)^2 
\left(\frac{L}{r}\right)^4+\dots
\\
&&
\nonumber
F_3(r,\theta,\varphi)=1+f_{33}(\theta,\varphi)\left(\frac{L}{r}\right)^3-\alpha^2  v_{0,\theta}^2(\theta,\varphi) \left(\frac{L}{r}\right)^4+\dots
\\
&&
\nonumber
F_0(r,\theta,\varphi)=1+f_{03}(\theta,\varphi)\left(\frac{L}{r}\right)^3+\left(-\frac{q^2}{L^2}+\alpha^2  v_{1}^2(\theta,\varphi)\right) \left(\frac{L}{r}\right)^4+\dots
\\
\nonumber
&&
S_1(r,\theta,\varphi)=\frac{1}{\sin\theta} \left( \sin\theta f_{23,\theta}(\theta,\varphi)+\cos \theta (f_{23}(\theta,\varphi)-f_{33}(\theta,\varphi))+s_{33,\varphi}(\theta,\varphi)  \right)
\log(\frac{L}{r})\left(\frac{L}{r}\right)^5+\dots
\\
\nonumber
&&
S_2(r,\theta,\varphi)=\frac{1}{\sin\theta} \left(  f_{33,\varphi}(\theta,\varphi)+\sin \theta  s_{33,\theta}(\theta,\varphi)+2\cos\theta s_{33}(\theta,\varphi)  \right)
\log(\frac{L}{r})\left(\frac{L}{r}\right)^5+\dots
\\
\nonumber
&&
S_3(r,\theta,\varphi)= s_{33}(\theta,\varphi)\left(\frac{L}{r}\right)^3+\alpha^2\frac{1}{\sin\theta}v_{0,\theta}(\theta,\varphi)v_{0,\varphi}(\theta,\varphi)\left(\frac{L}{r}\right)^4+\dots~.
\end{eqnarray}
In our approach, 
$v_0(\theta,\varphi)$ is imposed as a BC, while 
$v_1(\theta,\varphi)$ results from numerics.
The metric functions contain  
$f_{03}(\theta,\varphi)$,
$f_{23}(\theta,\varphi)$,
$f_{33}(\theta,\varphi)$,
$s_{33}(\theta,\varphi)$
which are also
fixed by the numerics,
subject to the constraints (which follow from the field eqs.):
\begin{eqnarray}
\nonumber
&&
f_{00}+f_{23}+f_{33}=0,
\\
&&
\label{conds} 
\cos\theta(f_{23}-f_{33})+s_{33,\varphi}+\sin\theta f_{23,\theta}+\frac{4}{3}\alpha^2\sin\theta v_1v_{0,\theta}=0,
\\
\nonumber
&&
2\cos\theta s_{23}-f_{33,\varphi} +\sin\theta s_{33,\theta}+ \frac{4}{3}\alpha^2\sin\theta v_1v_{0,\varphi}=0.
\end{eqnarray}

The mass  of the solutions is computed by employing the boundary counterterm approach in 
\cite{Balasubramanian:1999re}, as the conserved charge  associated with Killing symmetry
$\partial_t$ 
of the induced boundary metric,  found for a large value $r=$constant.
 A straightforward computation leads to the following expression:
\begin{eqnarray}
M=M^{(b)}+\frac{1}{8\pi G}\frac{3L}{2} 
\int_0^{2\pi} d\varphi  \int_0^\pi d\theta  \sin \theta
\left(
f_{23}(\theta,\varphi)+f_{33}(\theta,\varphi)
\right)~,
\end{eqnarray}
with
\begin{eqnarray}
 M^{(b)}=\frac{r_H }{2G}\left(1+\frac{r_{H}^2}{L^2}+\frac{q^2}{r_H^2} \right)
\end{eqnarray}
the contribution from the background metric.
Note that the same result can be derived by using the Ashtekar-Magnon-Das conformal mass
definition \cite{Ashtekar:1999jx}.

We expect these solutions to be relevant in the context of AdS/CFT and more generally
in the context of gauge/gravity dualities.
Thus it is of interest to evaluate the holographic stress tensor.
In order to extract it,
we first transform the (asymptotic metric) into Fefferman--Graham coordinates, by using a new radial coordinate $z$, with
\begin{eqnarray}
\label{cd1}
 r=\frac{L^2}{z}-\frac{2r_H^2+L^2}{4L^2}z+\frac{r_H^4+(q^2+r_H^2)L^2}{6r_HL^4}z^2.
\end{eqnarray}
 In these coordinates, the line element can be expanded around $z=0$  ($i.e.$ as $r\to \infty$) 
in the standard form
\begin{eqnarray}
\label{FG1}
ds^2=\frac{L^2}{z^2}
\bigg [
dz^2+\big(g_{(0)}+z^2 g_{(2)}+z^3 g_{(3)}+O(z^4)\big)_{ij}dx^i dx^j 
\bigg],
\end{eqnarray}
where $x^i=(\theta,\varphi,t)$ and 
\begin{eqnarray}
\label{FG2}
 \big(g_{(0)}+z^2 g_{(2)} \big)_{ij}dx^i dx^j =\left(L^2-\frac{z^2}{2}\right)(d\theta^2+\sin^2\theta d\varphi^2)-\left(1+\frac{z^2}{2L^2}\right)dt^2.
\end{eqnarray}
Then the background metric upon which the  dual field theory resides is
$d\sigma^2=g_{(0)ij}dx^i dx^j=-dt^2+L^2 (d\theta^2+\sin^2 \theta d\varphi^2)$, 
which corresponds to a  
static Einstein universe in $(2+1)$ dimensions.

From (\ref{FG1}) one can read the $v.e.v.$ of the holographic stress tensor \cite{deHaro:2000xn}:
\begin{eqnarray}
\label{FG3}
  \langle\tau_{ij}\rangle=\frac{3L^2}{16\pi G}  g_{(3)ij}=   \langle\tau_{ij}^{(0)}\rangle+  \langle\tau_{ij}^{(s)}\rangle,
\end{eqnarray}
being expressed as the sum of a background part plus a Maxwell contribution (which possesses a nontrivial
$(\theta,\varphi)-$dependence), 
\begin{eqnarray}
\label{Tij0}
&&
   \langle\tau_{ij}^{(0)}\rangle dx^idx^j=\frac{1}{16\pi G}\left(\frac{r_H^3}{L^2}+\frac{q^2+r_H^2}{r_H}\right)
	\left(d\theta^2+\sin^2\theta d\varphi^2+\frac{2}{L^2}dt^2 \right),
	\\
\label{Tijs}
	&&
	   \langle\tau_{ij}^{(s)}\rangle dx^idx^j=\frac{3}{16\pi G}\frac{1}{L^2} 
	\bigg( 
	f_{33}(\theta,\varphi)d\theta^2
	+2 s_{33} (\theta,\varphi)\sin \theta d\theta d\varphi
	\\
	\nonumber
		&&
		{~~~~~~~~~~~~~~~~~~~~~~~~~~~}	
		+f_{33}(\theta,\varphi)\sin^2\theta d\varphi^2
		-\frac{1}{L^2}f_{03}(\theta,\varphi)dt^2 
	\bigg).
\end{eqnarray}
 As expected, this stress-tensor is finite and traceless.
Also, it satisfies the conservation law in the presence of a 
background electric field
\begin{eqnarray}
  \langle\tau^{ab}\rangle_{;a}+j_a {\cal F}^{ab}=0,
\end{eqnarray}
a relation which holds via (\ref{conds}).
Here $j= v_1dt$ is the current density on the boundary 
and ${\cal F}=d {\cal A}$ (with ${\cal A}=v_0dt$) 
is the boundary electromagnetic field.

 \begin{small}
 
 \end{small}

\end{document}